\documentstyle[preprint,prl,aps]{revtex}
\begin{document}
\draft
\title{Weak first-order transition in the quasi-one-dimensional
         frustrated XY antiferromagnet}
\author{M.L. Plumer}
\address{Centre de Recherche en Physique du Solide et D\'epartement de
Physique}
\address{Universit\'e de Sherbrooke, Sherbrooke, Qu\'ebec, Canada J1K 2R1}
\author{A. Mailhot}
\address{INRS-EAU, 2800 rue Einstein, C.P. 7500, 
              Ste.-Foy, Qu\'ebec, Canada G1V 4C7}
\date{May 1996}
\maketitle
\begin{abstract}  The results of finite-size scaling analysis of
histogram Monte-Carlo simulations on the stacked triangular XY
antiferromagnet with anisotropic near-neighbor exchange interactions
are presented.  With ferromagnetic interplane coupling ten times stronger than
the antiferromagnetic intraplane interaction ($J_\| / J_\bot = -10$), a weak 
first-order transition
is revealed.  These results represents the first simulational corroboration
of a wide variety of renormalization-group calculations made over the past
twenty years.  As such, they shed light on recent controversy regarding
the critical behavior in this and similar frustrated systems and have
particular relevance to recently reported data on $CsCuCl_3$.

\end{abstract}
Although the earliest, as well as the most recent, efforts using $4-\epsilon$
renormalization-group (RG) techniques
to study the critical behvior in simple frustrated antiferromagnets (AF)
indicated that these systems undergo a fluctuation-induced 
first order transition, a
plethora of other recent studies have resulted in a wide variety of 
scenarios.  The recent interest in this field has been inspired by the proposal
of Kawamura of new {\it chiral} universality classes in the XY and
Heisenberg cases based on his study of the $4-\epsilon$ RG expansion as well
as Monte Carlo (MC) simulations \cite{kawa1}.  This suggestion 
is in contrast with
results from a $2+\epsilon$ RG expansion of the nonlinear sigma model
($NL \sigma$) from which either an $O(4)$ (in the Heisenberg case), tricritical,
or first-order, transition is expected \cite{azaria}.  It is important to note
that most of the critical exponents estimated by Kawamura are not so
different from those expected at a mean-field tricritical transition.
Although all subsequent MC work has been in support of Kawamura's
chiral universality for the Heisenberg system \cite{mail}, nearly two years 
ago the present authors published
the results of extensive MC simulations which gave strong
support to the idea of tricriticality in the $XY$ case \cite{plum1}. 
In that work (hereafter referred to as I), a simple hexagonal lattice
was considered with AF exchange interactions in the triangular
plane $J_\bot = 1$, giving rise to frustration, and ferromagnetic coupling 
along the $c$-axis, $J_\| = -1$.  This model thus represents isotropic
near-neighbor couplings.  It was also studied with the addition
of an in-plane magnetic field $H$, where a weak first order transition
was revealed to an unfrustated 3-state Potts phase at small values of 
H \cite{plum2} (hereafter referred to as II).  We report here MC simulation
results for the $XY$ model using $J_\bot = 1$ and $J_\| = -10$,
which represents somewhat quasi-one-dimensional exchange anisotropy.
Finite-size scaling of these new data show strong evidence of
a first-order transition {\it at} the ordering temperature but with a tendancy 
toward tricritical behavior at a slightly higher temperature.  These
results are consistent with very recent specific heat data
on $CsCuCl_3$ \cite{weber}.

Relevant work published prior to 1995 is discussed in I, Ref. \cite{plum3},
and references therein. It is useful to review here the more recent results
pertaining to the present study, beginning with those based on standard
RG theory.  Higher-order expansions in $4-\epsilon$
have reaffirmed results from the earlier studies in support of a first-order 
transition in both XY and Heisenberg cases \cite{anton}.  In contrast, 
a partial resummation based on the 1/N expansion has been shown
to yield a continuous transition for both models \cite{joli}.  
The delicate nature of this model is further revealed by the very recent 
proposal of yet another new fixed point based on RG calculations \cite{dobry}.
Detailed re-examinations of the $NL\sigma$ model have also given rise
to a variety of scenarios.  Azaria {\it et al.} \cite{aza2} have demonstrated
the existence of an additional length scale which may be relevant for the
interpretation of MC results on finite-size systems.  In another study,
this model has been shown to yield a variety of possibilities, 
including a two-step process (first proposed by Chubukov 
\cite{chub}) where a first order transition to a ``nematic" phase
is followed by a continuous transition to a chiral state as the
temperature is lowered \cite{dav}.   Although these results may be of
relevance to the present work (see below),
it should be emphasized that the validity of the $NL\sigma$ model 
of frustrated systems has been questioned \cite{zum} as, indeed, has the
standard Ginsburg-Landau-Wilson approach \cite{azaria,aza2}.    
Other recent theoretical work includes the observation of a two-step
transition process in a frustrtated two-dimensional model by MC simulations 
\cite{ols}, criticality controlled by a Lifshitz point in helimagnets
\cite{barb}, as well as the possibility to investigate chirality using
polarized neutrons \cite{mal}.

In addition to some recent review articles which summarize experimental
estimates of critical exponents \cite{loh}, several new results on
specific materials have appeared \cite {kato,dup,weber}.  Although
many of these data support the existence of chiral universality, 
most do so only marginally and some data clearly suggest $O(4)$
criticality or a first order transition.  Of particular relevance to 
the present work are the results on $CsCuCl_3$ \cite{weber}, discussed in
more detail below. 

As in I and II, the standard Metropolis MC algorithm was used in this work,
along with analysis based on the histogram technique, on 
$L \times L \times L$ lattices with L=12-33.
However, because the exchange anisotropy causes a lower Metropolis 
acceptance rate, longer runs (typically by about a factor of two)
were employed here, with thermodynamic averages per run calculated
using $1 \times 10^6$ MC steps per
spin (MCS) for the smaller lattices and $2.5 \times 10^6$ MCS 
for the larger lattices, after discarding the initial
$2 \times 10^5$ -- $7 \times 10^5$ MCS for thermalization.
Averaging was then made over 6 (smaller L) to 
16 (larger L) runs, giving $4 \times 10^7$ MCS for the calculation of 
averages with the larger lattices. Errors were then estimated by taking 
the standard deviation of results among the runs.  

A rough determination of the transition temperature was first made by 
examination of various thermodynamic quantities in short-run temperature
sweeps at $L=24$.  From these results, the estimate $T_c \simeq 4.6$
was made.  Histograms were then generated at $T_0=4.55, 4.59,
4.65, 4.69$ and 4.71 in an effort to determine $T_c$ more accurately.
Two methods were used.  At smaller lattice sizes ($L=12-27$), finite-size 
scaling of extrema in the suceptibility $\chi'$, as well as the logarithimic
derivative of the order parameter $V_1$ (see Ref. \cite{mail} for definitions),
were made for both spin and chiral order parameters.  Scaling
made with the assumption of tricriticality
($\nu = \frac12$) gives a reasonably good straight-line fit
and yields the estimate $T_c \simeq 4.67$.  The 
corresponding results for chiral order were nearly identical.  Scaling with 
the assumption of a first-order transition did not yield a good
straight-line fit. An accurate determination of the temperature at 
which extrema occur requires data from many histograms generated at different 
temperatures $T_0$.  Our data do not have sufficient precision for the
simultaneous determination of both $T_c$ and $\nu$.   

For this reason, we also applied the cumulant-crossing method to estimate the
critical temperatures corresponding to both spin and chiral orderings.
The results shown in Fig. 1 also reveal the effects of relativley large 
fluctuations and are not amenable to further finite-size scaling analysis 
(as performed in I) in an effort to better estimate $T_c$.  However, it may be 
observed from these data that 
the inverse critical temperature $\beta_c$ associated
with the spin order appears to be near 0.214 ($T_c \simeq 4.673$) 
whereas it is closer to 0.213 ($T_c \simeq 4.695$) for the chiral order.
It is not possible to claim that there are two distinct transitions by these 
data alone.  

Finite-size scaling of thermodynamic functions was performed at
the two temperatures T=4.673 and T=4.695.  The analysis is complicated
by the relatively large finite-size and fluctuation effects associated with
exchange anisotropy.  In an effort to allow for the possibility that
the smaller lattices used in this simulation were not sufficiently 
large to accomodate the true critical behavior, scaling of thermodynamic
quatities with the functional form $F_1 = a + b L^x$ was considered
in addition to the usual assumption $F_2 = b L^x$. (Of course, only
the form $F_1$ was used for the specific heat).  Differences between
the estimated critical exponents using these two fitting functions
varied considerably.  This is believed to be a reflection of the 
different relative strengths of finite-size effects depending on the
thermodynamic quantity under consideration \cite{pec}.  In some cases, strong
effects were also found if the smaller values of $L$ were not included
in the fit. 

In general, greater fitting-procedure effects were found
for the scaling at the lower temperature, $T=4.673$. 
Exponent estimates were generally larger if the form $F_1$ was used
and increased (in most cases) if the smaller $L$ data were excluded.
For example, in the case of the specific heat $C$
(where $x = \alpha / \nu$), exponent values 1.3(3), 1.7(5), and 2.5(10)
(where errors reflect only the robustness of the fit) were found
using only data for $L$=18-33, 21-33, and 24-33, respectively.
The dependence of exponent estimates on the assumed fitting function, as well
as excluded data, are presented in Table I for $\chi'$ 
($x = \gamma / \nu$), $V_1$ ($x = 1 / \nu$), and the order parameter $M$
($x=-\beta / \nu$).  (Only results of fitting the form $F_2$ to $M$ are
presented due to the large errors found if the form $F_1$ is assumed). 
The results of these fits are very suggestive of a first-order transition.
Note that although no discontinuity is observed in $M(T)$ at $T_c$, 
the small exponent value is indicative of a very sharp rise near
the transition, as expected if a jump in $M$ is
smoothed-out due to significant finite-size effects.

Confirmation that finite-size scaling at $T=4.673$ indicates a first-order
transition is demonstrated by Figs. 2-4 where good asymptotic straight-line
fits are found with the assumption that $x=3$ for $C$, $\chi'$ and
$V_1$ for both spin and chiral order.  {\it The data presented in 
this manner are convincing in view of the discussion in Ref. \cite{pec},
and in particular, of the similar scaling found in II
for the transition to the 3-state Potts phase}.  Although it is
known from rigorous symmetry arguments that the 3-state Potts transition
is indeed first order, this is revealed only at the larger lattice
sizes in MC simulations and is indicative of a very small latent
heat. The observation of only asymptotic volume dependence at
a weak first order transition has been previously emphasized
\cite{pec}. Similar conclusions can be  made regarding the present model.
That the transition here is indeed only very weakly first order, is  
also evident by the observation of a single peak in the
energy histograms as well as our estimate for the energy
cumulant (see I and II), $U^\ast = 0.666~6631(3)$.  In addition, the assumption
of tricritical exponents yields scaling as in Figs. 2-4 with a
noticable (but small) inferior quality based on goodness-of-fit ($R^2$)
tests.

In contrast to these results, finite-size scaling at the higher
temperature $T=4.695$ yields exponents closer to those expected
of tricritical behavior (and with diminished effects due to the 
smaller lattices), where 
$\alpha / \nu = 1$, $\beta / \nu = \frac12$, $\gamma / \nu = 2$, 
and $1 / \nu = 2$.   
For the specific heat, exponent values 0.3(3), 0.6(7), and 0.9(12)
were found using using $L$=18-33, 21-33, and 24-33, respectively.
Corresponding results for $\chi'$, $V_1$ and $M$ are presented in Table I.
Scaling with the assumption of tricritical exponents yields 
a similar qualitity of asymptotic straight-line fit to the data as
presented in Figs. 2-4, whereas the assumption of volume dependence
yields a somewhat (but clearly) inferior fit.

With the assumption that the transition temperature of the present model is 
close to 4.67 (based on the reasonable premise that the spin (and not
chiral) order parameter is more relevant in determining $T_c$),
the MC analysis presented here is strongly
supportive of several of the very recent theories and experimental
results associated with phase transitions in geometrically frustrated
systems.  The fact that we find a stronger tendancy towards a
first-order transition in the present somewhat-quasi-one-dimensional
exchange model than in I is consistent with arguments put forth in I. 
In that work, it was noted that the proximity of the 3-state Potts
phase could generate an effective cubic term in the Hamiltonian,
the relative importance of which increases with increasing
short-range order along the c-axis chains.  Such short-range order
is enhanced by a larger value of $J_\|$.  These effects are of particular
importance in the present system since the three-dimensional ordering
temperature is reduced by frustration allowing short-range order
to be well developed by the time $T$ is reduced to $T_c$.  The very recent
mean-field treatment of the $NL\sigma$ model in Ref. \cite{dav} 
also indicates the importance of cubic contributions to the 
Hamiltonian.

Perhaps the recent theoretical work of Azaria {\it et al.} \cite{aza2} is most
relevant.  Their conclusion that true critical behavior 
may be revealed with MC simulations only by using relativley large 
lattices is fully consistent with the present results and may
eventually be proven relevant for the Heisenberg model.

Conclusion regarding the critical behavior in the
somewhat-quasi-one-dimensional ($J_\|  / J_\bot \simeq -5$)
compound $CsCuCl_3$ also appear
to support the scenario of a very weak first order transition for 
frustrated $XY$ systems \cite{weber}.  These specific-heat data indicate
an unusually large exponent $\alpha \simeq 0.35$, suggestive of tricritical 
behavior, except very close to the transition where
first-order behavior is observed. A similar scenario has
also been put forth based on the analysis of experimental data
on several rare-earth helimagnets \cite{dup}.  

In conclusion, the finite-size scaling analysis presented here is the
first evidence from MC simulations that the frustrated hexagonal
antiferromagnet exhibits a weak first order phase transition to 
the paramagnetic state.  This work serves to strengthen the conclusions
of several older, and also the most recent, of $4-\epsilon$ RG analyses
as well as new studies based on the $NL\sigma$ model.  In addition, 
a clearer picture is emerging (proximity of the 3-state Potts phase)
that explains the difficulty in treating this long-standing problem 
theoretically, experimentally and numerically.

\acknowledgements
We thank E. Sorensen for suggesting this work, U. Schotte for discussions on 
$CsCuCl_3$,
and the Service de l'Informatique for the partial use of 10 RISC 6000
Workstations.  This work was supported by NSERC of 
Canada and FCAR du Qu\'ebec.
%

\begin{table}
\caption{ Dependence of exponents x on the assumed
critical temperature, fitting function (see text) and excluded data.
Errors indicate only the robustness of the fit.}
\vskip0.2cm
\centerline{$T=4.673$} 
\begin{tabular}{cccccccc}   
& & $F_1$ & & & &  $F_2$  &\\
\tableline
L =              & 15-33 & 18-33  & 21-33   & &  15-33 & 18-33  & 21-33 \\
\hline
$\chi'$          & 2.9(1)  & 2.7(1) & 2.8(1)& & 2.36(4)& 2.38(4)& 2.41(4)  \\ 
$V_1  $          & 2.3(1)  & 2.5(2) & 2.7(3)& & 2.00(5)& 2.02(6)& 2.07(6)  \\
$M    $          &         &        &       & &-0.365(5)&-0.364(8)&-0.347(3)  \\
\end{tabular} 
\vglue0.3cm
\centerline{$T=4.695$} 
\begin{tabular}{cccccccc}   
$\chi'$          & 2.17(6)  & 2.3(1) & 2.2(2)& & 2.12(2)& 2.12(2)& 2.13(2)  \\
$V_1  $          & 1.83(7)  & 1.9(1) & 1.9(2)& & 1.78(1)& 1.78(2) & 1.79(2)  \\
$M    $          &         &        &       & &-0.48(1) &-0.501(7)&-0.51(1)  \\ 
\end{tabular} 
\end{table}

\begin{figure}
\caption{(a) Spin order-parameter cumulant crossing as a
function of inverse temperature ($\beta = 1/T$) for the lattice
sizes indicated.  (b) Corresponding results for chiral order.}
\label{fig1}
\end{figure}

\begin{figure}
\caption{Finite-size scaling at $T=4.673$ 
of the specific heat data for $L=12-33$.
Data at L=12-21 are excluded from the fit.  Error bars are estimated
from the standard deviation found in the MC runs.}
\label{fig2}
\end{figure}

\begin{figure}
\caption{Finite-size scaling at $T=4.673$ of the spin susceptibility $\chi'$
as well as the logarithmic derivative of the order
parameter $V_1$ (see text).  Data at L=12-18 are excluded form the fit.}
\label{fig3}
\end{figure}

\begin{figure}
\caption{Thermodynamic functions associated with chiral order, as in Fig. 3.}
\label{fig4}
\end{figure}

\end{document}